# Amortized Clustering Assistant Classification of Anomalous Hybrid Floquet Modes in a Periodically Driven non-Hermitian Lattice

Yifei Xia, [1] Xiumei Wang, [2*] Yali Li,[3] and Xingping Zhou, [1*]

[1] *Institute of Quantum Information and Technology, Nanjing University of Posts and Telecommunications, Nanjing 210003, China*

[2] *College of Electronic and Optical Engineering, Nanjing University of Posts and Telecommunications, Nanjing 210003, China*

[3] *School of Foreign Studies, Nanjing University of Posts and Telecommunications, Nanjing 210003, China*

*wxm@njupt.edu.cn

*zxp@njupt.edu.cn

The interplay between Floquet periodically driving and non-Hermiticity could bring about intriguing novel phenomena with anomalous Floquet topological phases of a finite-size, tight-binding lattice model. How to efficiently investigate on quasi-energy and eigenfield of a non-Hermitian Floquet system with complicated driving protocol remains a challenging task. In this work, we define a somewhat complex driving protocol for a bipartite lattice system and discover two nontrivial topological phases that support Floquet $\pi$ mode. Thereafter, we introduce unsupervised learning method in order to explore distribution features of system eigenfunctions under different magnitude of system energy gain/loss. We utilize the idea of amortized clustering and construct an algorithm selector that could dynamically upgrade with increasing gain/loss as input parameter. Proper employment of the selector enables us to reveal the regulation of dynamic localization from abundant possible wave function distribution in two-dimension lattice in another efficient way. In addition, our work provides a feasible methodology via machine learning method to assist in classification of Floquet modes.

# I. Introduction

Over the past decades, Floquet systems with time-periodic Hamiltonians have attracted great attention and played significant roles in the research of quantum dynamics under external driving fields [1-5]. Time-periodic driving has become more prevalent lately to engineer Floquet topological matter [6-10]. Under the irradiation of electromagnetic fields, semimetal and normal insulators can be driven into nontrivial phases [7, 11], which may support novel topological states, such as Floquet winding metals [12-14], anomalous Floquet topological insulators of first or higher order [15-18]. Meanwhile, recently introducing non-Hermiticity into open Floquet systems has become an efficient method. With complex energies, non-Hermitian systems may support richer topological structures which differ from those in Hermitian systems [19-21]. Exceptional points (EPs) [22-25] and non-Hermitian Skin Effect (NHSEs) [26-28] are two prominent phenomena exhibiting unique features of non-Hermitian systems.

Combing Floquet engineering and non-Hermiticity, the research of open systems receives a great extension [29-33, 39]. The recent works have discovered numerous intriguing features of non-Hermitian Floquet topological system such as Floquet EPs and Floquet NHSEs [34, 35]. With further in-depth research, multiple topological transitions and spectral singularities become novel exhilarating discovery [20, 36-39]. By revising the driving protocol of Floquet engineering, new topological phenomena could be discovered without the change of physical lattice model. With the complexity level of non-Hermicity and Floquet driving protocol increasing, the system may possess more complicated eigenstates, meaning that manually finding the features of eigen functions becomes rather time-consuming and difficult. Therefore, introducing machine learning and utilizing intelligent clustering algorithms may become a bright resolution [43-45]. The construction of an intelligent algorithm selector is capable of efficiently classifying typical topological modes and discovering new eigenfunction distribution, hence it can serve as a useful tool for promoting and testing theorical analysis.

In this work, we introduce more freedoms concerning non-Hermicity and a special driving protocol for a non-Hermitian Floquet bipartite lattice system, and thus discover two singular topological phases with peculiar complex quasi-energy. It is proved that the equivalence of the impact that gain/loss and nonreciprocity exert on the system. Later, we set the gain/loss as major variable and observe the quasi-energy bands, to find its nontrivial property of dynamic stabilization and localization. In a bid to further explore the distinctive features of new topological phases, we introduce machine learning method, draw on the experience of the idea of meta-learning [46-49] and construct an algorithm selector model, with a library containing carefully selected algorithms like agglomerative clustering and mean-shift. After training the model and applying the core amortized clustering method [46, 47], the singular eigenfunction distribution supporting hybrid corner-edge mode and edge-bulk mode is disclosed, which corroborates theoretical analysis well. As we gradually build up an algorithm learner and solver, we also provide a preliminary train of thought utilizing unsupervised learning to address similar problems.

This organization of paper is shown as follows. In Sec. II, we introduce our basic model and analyze its quasi-energy evolution features. In Sec. III, by slightly changing the Floquet driving protocol, we establish a fine-tuned model, revealing its singular Floquet bands as well as two topological transition phases. In Sec. IV, unsupervised learning method is introduced in a bid to help clarify numerous eigenfunctions of fine-tuned model introduced in Sec. III with different model parameters. We draw the idea of meta-learning and amortized clustering and construct an intelligent algorithm selector model, which is then tested by basic model and utilized to analyze fine-tuned model. Ultimately, under certain gain/loss parameters, the model can output clusters with singular eigenfield that characterized hybrid topological mode, illustrating dynamic stabilization and localization features of the two topological phases supporting Floquet $\pi$ mode. In Sec. V, we discuss the strengths and weaknesses of our meta-learning model and consider our further explosive work.

# II. Basic model

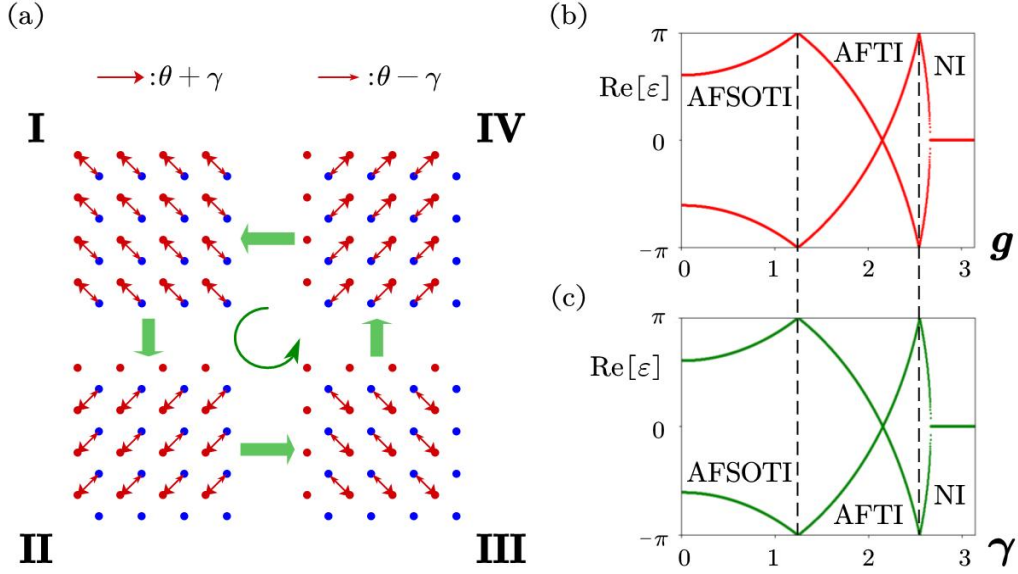

FIG. 1. Basic model of two-dimensional Floquet bipartite lattices and its bandwidth evolution as a function of two non-Hermitian factors. (a) The construction of the model, which runs under a four-step driving protocol. The couplings in each step are fully dimerized yet nonreciprocal. The larger arrow tip represents the stronger coupling θ + γ, while the smaller one the weaker coupling θ − γ. Each site is sequentially coupled with one of the four neighboring sites in a counterclockwise direction. Gain and loss sites are alternatively distributed, with red site representing gain and blue site representing loss. (b) and (c) are respectively the evolution of real part of model's quasi-energy with gain/loss parameter $g$ and nonreciprocity parameter γ.

Our basic model originates from Ref. [36], which is a time-dependent, tight-binding lattice, with driving protocol illustrated in Fig. 1(a). The entire driving protocol is divided into four period with equal interval. Each period involves dimerized inter-site coupling, with each site sequentially coupled with one of its four neighboring sites in a counterclockwise direction. Every unit cell has two sublattices with balanced gain and loss, whose magnitude is denoted as a variable $g$. Additionally, we introduce another parameter γ to characterize the strength of nonreciprocity. Under the periodic boundary condition (PBC), the time-dependent Hamiltonian in momentum space in four period can be written by

$$H(k,t) = H_n(k), lT + (n-1)T/4 < t < lT + nT/4, \quad (1)$$

where $l \in Z$ and n = 1, 2, 3, 4, meaning the index of four different period. T is the entire driving period and can be simply set as T = 4. $H_n(k)$ is given by

$$H_n(k) = (\theta+\gamma)e^{ib_n k}\sigma_+ + (\theta-\gamma)h.c. + ig\sigma_z. \quad (2)$$

Here, θ is the basic coupling parameter, while γ is another coupling parameter representing the strength of nonreciprocity. $\sigma_{x,y,z}$ are Pauli matrices, and $\sigma_\pm = (\sigma_x \pm i\sigma_y)/2$. The vectors $b_n$ are given by $b_1 = (0,0), b_2 = (0,1), b_3 = (-1,1), b_4 = (-1,0)$. We concentrate on the ground state of the system and just set Bloch wavevector $k$ as zero vector [40]. Then, actually for any moment $H_n(k)$ is given by

$$H_n(k) = H = \begin{pmatrix} ig & \theta+\gamma \\ \theta-\gamma & -ig \end{pmatrix}. \quad (3)$$

In a bid to obtain quasi-energy dispersion $\varepsilon(k)$, it is a necessity to solve the Floquet eigenvalue equation given by

$$U_T(k)|\Psi(k)\rangle = e^{-i\varepsilon(k)}|\Psi(k)\rangle, \quad (4)$$

where $U_T(k) \equiv \Im\exp\left[-i\int_0^T H(k,\tau)d\tau\right]$ is the Floquet operator. Here, $\Im$ is the time-ordering operator.

With the increase of the *g*, multiple topological transitions appear, which are shown in Fig. 1(b). We set the coupling parameter θ as 0.9π and nonreciprocity parameter γ as 0.3π, meaning that in the absence of gain and loss, the Hamiltonian of system supports anomalous Floquet second order topological insulator (AFSOTI). As the variable *g* increases, the topological phase gradually turns into anomalous Floquet topological insulator (AFTI) and normal insulator (NI).

The two parameters *g* and γ determine the strength of non-Hermiticity of Floquet system. We then set *g* as a constant 0.3π and increase γ from 0 to π, meaning that we swap the setting of the two parameters. Similar result is shown in Fig. 1(c), with almost the same curve and trend. Solving the Eq. (4), we obtain the quasi-energy evolution with time:

$$\varepsilon(t) = \sqrt{\theta^2 - g^2 - \gamma^2}\, t \mod (-\pi, \pi]. \quad (5)$$

The solution above demonstrated the equivalence of $g$ and $\gamma$, as they both equally determine the linear growth trend of $\varepsilon$. As a result, gain/loss and nonreciprocity actually equally impact on the non-Hermiticity of the Floquet system. Hence, in the following research, we fix $\gamma$ as $0.3\pi$ and regard $g$ as the sole variable.

## III. Fine-tuned model

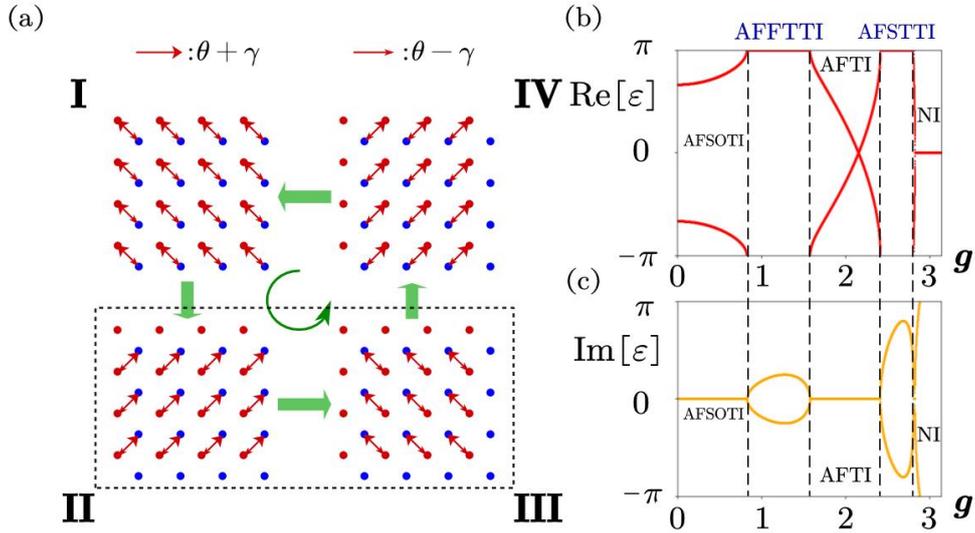

FIG. 2. (a) The construction of fine-tuned model. The differences can be seen in period II and III. In the fine-tuned model, the directions with stronger coupling are opposite to original model in the two periods, which brings about novel properties of the system. (b) Bandwidth evolution as a function of $g$ for the fine-tuned model. Compared to the common model, the function has two intervals with consistent function value $\pm\pi$, which may support novel topological phases of the system. The demarcation points of five different topological phases are approximately at $g = 0.83$, 1.58, 2.41 and 2.81. The novel topological phase zones are slightly colored grey. (c) Imaginary part of quasi-energy evolution as a function of $g$ for the fine-tuned model. When the real part reaches $\pm\pi$, the imaginary part synchronously becomes nonzero, indicating that the stability of system decreases. The two novel transition phases both possess nonzero imaginary part of Floquet

system quasi-energy.

In order to further explore the influence of non-Hermiticity, we build a fine-tuned model and slightly change its driving protocol, which merely affect the coupling in period II and III. The changed driving protocol is given by

$$H_n(k) = \begin{cases} (\theta+\gamma)e^{ib_n k}\sigma_+ + (\theta-\gamma)h.c. + ig\sigma_z, \text{ for period } I, IV, \\ (\theta-\gamma)e^{ib_n k}\sigma_+ + (\theta+\gamma)h.c. + ig\sigma_z, \text{ for period } II, III. \end{cases}$$

In the original model, the stronger coupling $\theta + \gamma$ always points to the red site, namely the gain site. In the fine-tuned model, the stronger coupling tends to point upwards, making the whole system appears like undergoing a directional flow of energy. The time-dependent Hamiltonian becomes more complicated, and bring about especial quasi-energy evolution with *g*, as shown in Fig. 2(b). Now the Hamiltonian may support five kinds of topological phase, with two kinds of transition phase that possess $\pm\pi$ quasi-energy. The corresponding insulators are named as anomalous Floquet first/second transition topological insulator (AFFTTI/AFSTTI).

The solution of Eq. (4) becomes more complicated, and the quasi-energy evolution with time is no longer linear. The comparation between two models are shown below in Fig. 3. Quasi-energy turns to be complex values for *g* in some certain ranges, meaning that there is energy dissipation in the whole system. With gain/loss as well as directional energy flow caused by nonreciprocal coupling, the Floquet system shows more intricate properties, and the multiple topological transition becomes even more frequent.

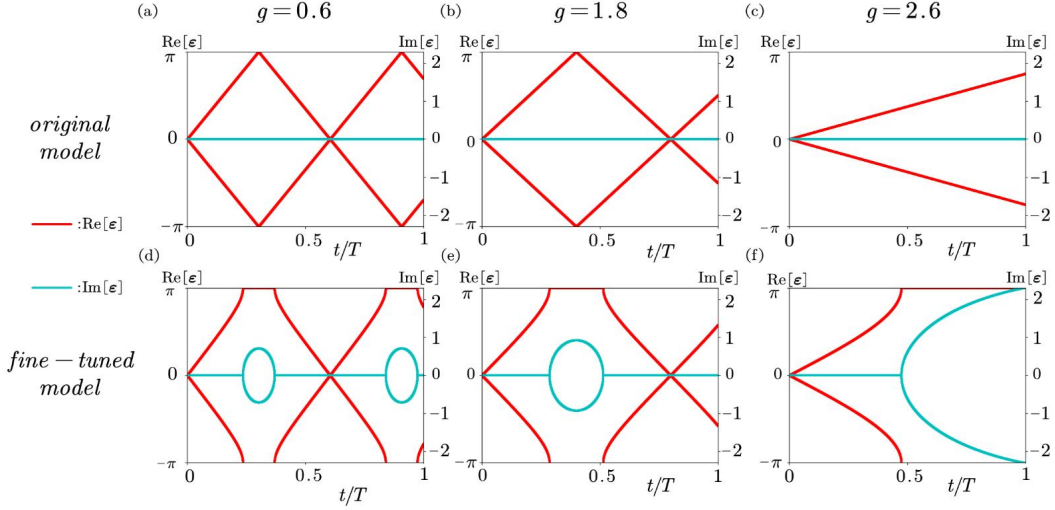

FIG. 3. Comparison between quasi-energy evolution with time of basic and fine-tuned models. (a)-(c) correspond to original model. The blue curve represents real part, and purple curve represents imaginary part. (d)-(e) correspond to fine-tuned model. The red curve represents real part, and cyan curve represents imaginary part. To vertically compare, (a) (d) (g) share the same gain/loss 0.6, and (b) (e) (h) 1.8, (c) (f) (i) 2.6. In the original model, the imaginary part of quasi-energy is constantly zero. In the fine-tuned model, the band inversion no longer happens at a certain point but a transition range with the value of quasi-energy equal to ±π, and the imaginary part is nonzero in the range.

It is easily to discover the existence of Floquet π mode in Fig. 3, with consistent quasi-energy gap and nonzero imaginary part [31], which may induce another kind of singular topological mode with dynamic localization [41]. The system will enter a dynamic stable state probably caused by special driving protocol that regulate energy flow with certain direction, and the nontrivial state will disappear as greater gain/loss repress this effect. Before NI, such kind of dynamic stable state appears for twice, which might imply the law of system dynamic stabilization evolution with magnitude of non-Hermicity.

For the original basic model, the presence of various topological modes including corner mode and edge mode can be further confirmed by studying the quasi-energy bands under open boundary condition (OBC), meaning that the both *x* and *y* directions of the lattice are length-finite. Just as shown in Fig. 3, we set the finite size as N = 8, so the entire system has a finite number of 64 sites, with 32 gain sites and 32 loss sites which are diagonally staggered arranged. Nevertheless, for our fine-tuned model, the proof offered by quasi-energy bands and that can confirm the presence of singular topological modes gets more difficult to obtain, and we are not clear about whether those trivial modes or nontrivial modes exist or not in advance.

## IV. Further Research by Machine Learning Method

We further investigate the quasi-energy bands as well as their respective eigenfunctions, in order to reveal the law of eigenfunction distribution that dynamic stabilization and localization find expression in. However, conventional mathematical and physical solutions, such as Fourier series expansion and extended Hilbert Space method [42], become rather complicated and time-consuming on account of the increased complexity of Floquet driving protocol. Hence, in the following research, we introduce unsupervised learning method to assist us in classifying the huge number of eigenfunctions, aiming to find out the law of topological modes and meanwhile try to establish a reliable methodology that do well in similar problems.

The overall assignment is defined as a clustering task, which requires a number of datasets and a few clustering algorithms. Apparently, each single clustering algorithm may not be suitable for all situations. In comparison with trying multiple algorithms one by one, building an intelligent algorithm selector is obviously a wise scheme, enabling us to select the most efficient algorithm in accordance with parameter input. Therefore, we utilize the idea of meta-learning for reference [47, 48], and construct the meta-learning framework and flowchart which is shown as Fig. 4. We fix some parameters of Floquet non-Hermitian system as size N = 8, strength of coupling $\theta = 0.9\pi$ and nonreciprocity $\gamma = 0.3\pi$, and choose the magnitude of gain/loss $g$ as the sole variable, to wit the input parameter for the task extraction module, where each input parameter $g$ can correspond to a dataset with $2 \times N \times N = 128$ eigenvectors, then it will be a basic task for the algorithm selector. The emerging idea of meta-learning helps build the machine learning framework and clarify optimization goals.

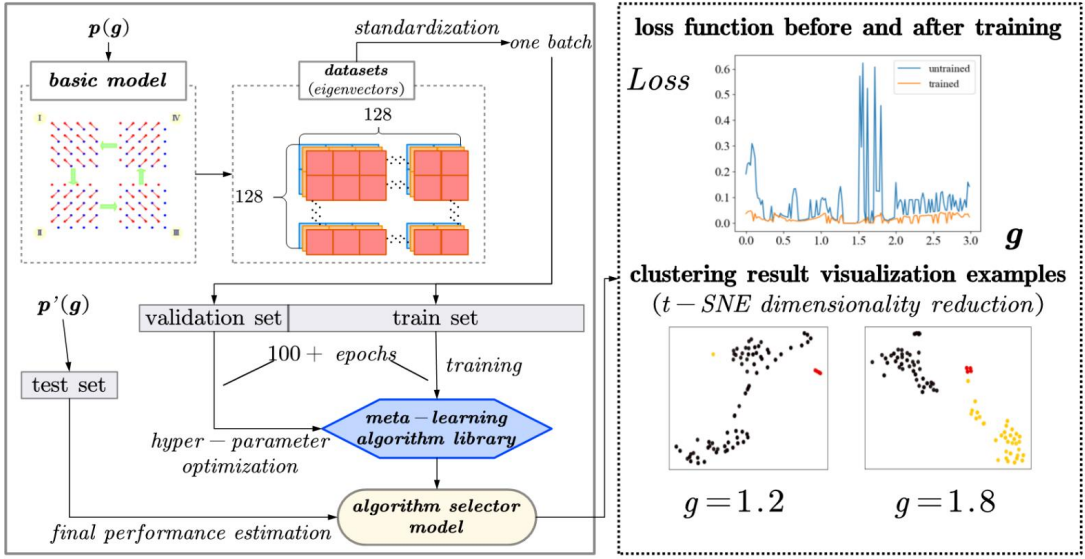

FIG. 4. Flowchart of the meta-learning training course. The $p(g)$ is the distribution set of gain/loss, the input used to formulate train set and validation set. Another distribution of gain and loss is used to formulate the test set that plays role in final performance estimation. The right part roughly displays the final outcomes of trained selector for the specific physical non-Hermitian Floquet system.

The magnitude of gain/loss $g$ serves as the physical parameter input and also the mere tag of each basic clustering task. When training the meta-learning model, we sample the value of $g$ from range [0, 3] with equal stride = 0.01, which derives from rule of thumb. The model will traverse each algorithm from library and invoke the clustering solver when processing each basic task. The final goal is to make the model learn 'how to cluster', i.e., when receiving new clustering tasks, the model is still able to select the algorithm with the best performance. The algorithms also require hyper-parameter optimization, where we adopt grid research method (seen in Appendix B) to obtain the optimal solution.

A straightforward method to measure the clustering performance is to draw the t-SNE distribution image containing clustering point labels (seen in Appendix C). In a bid to measure the performance, it is a necessity to define the 'loss function' [50, 51] for the model. Since we are going to evaluate the effectiveness of clustering outcome, a classical cost index for clustering evaluation, Davies-Bouldin Index (DBI) [52, 53], is used to form the 'loss function'. Its core idea is to judge the clustering effect by quantifying the ratio of the tightness of the samples within the cluster to the separation between the clusters. DBI can be calculated as

$$DBI = \frac{1}{n}\sum_{i=1}^{n}\max_{j\neq i}(\frac{S_i + S_j}{d_{ij}}), \tag{7}$$

where $S_i, S_j$ represent the intra-cluster dispersion of cluster $i, j$, and $d_{ij}$ is the distance between cluster centers. DBI is a cost index, meaning that the smaller value of DBI, the better clustering results. A small DBI value means that the sample points within the cluster are tightly packed, and the distance between clusters is large. Thus, DBI can be a vital component for the loss function.

Similarly, Silhouette Coefficient (SC) is also a kind of evaluation index that focuses on the similarity property, and it is a widely-used metric for evaluating the quality of clustering results [54]. It combines the concepts of cohesion within clusters and separation between clusters. The coefficient is calculated for each data point, taking into account two distances: the average distance from the point to all other points in its own cluster ($a_i$), and the average distance from the point to all points in the nearest neighboring cluster ($b_i$). SC for a data point $i$ is defined as

$$SC = \frac{b_i - a_i}{\max(a_i, b_i)}. \tag{8}$$

The overall SC for a clustering is the mean of the coefficients for all data points, with value range [-1,1]. A higher value indicates better clustering, where data points are well-matched to their own cluster and poorly matched to neighboring clusters. Considering SC may be negative, the entire loss function is defined as

$$L(g_i) = \frac{DBI_i}{SC_i + 1} pt, \tag{9}$$

where $pt$ is the operation time factor, which is defined as

$$pt = 1 + \lg(t)/10, \tag{10}$$

where *t* is the operation time. We define the time factor $pt$ as a cost index in order to balance the influence given by clustering results, yet the clustering effectiveness is still the vital consideration. If the fast algorithm possesses poor clustering performance, it will still get a prominently high value of loss function, meaning that it is not suitable for the specific clustering task.

Loss function can serve as the core component of the optimization goal [50]. With the definition of loss function, the final optimization goal can be defined as

$$\min E_{g_i \sim p(g_i)}\{L[F(g_i)]\},$$

where $g_i$ is the magnitude of gain/loss and the mere tag for each task, and $p(g_i)$ stands for the distribution of tasks, taking the uneven lengths of different topological ranges into consideration. $F$ is the trained algorithm selector, and $L$ is the loss function. The optimization goal indicates that a well-trained algorithm selector can minimize the mean value of all outcomes of loss function.

The one traversal of sampling points can be perceived as a single batch, and we carry on over 100 epochs in order to acquire better hyper-parameters as well as a better mapping relationship between algorithms and clustering tasks. Details for the training course can be seen in Fig. 4, and its right section roughly displayed an algorithm selection plan for the basic Floquet model of different topological phases and a few clustering results visualization images.

After training the algorithm selector, we first test it on the basic model, whose eigenfunction distribution is established and simple. The algorithm selector program can output indices of eigenfunction that each cluster contains. We draw the corresponding eigenfields and reveal the distribution features of corner mode, edge mode and bulk mode. The different localized corners will be classified as multiple clusters, yet they can be easily further classified manually. We also calculate the value of SC and DBI/SC for the selected and unselected algorithms as illustrated below in Fig. 5. It indicates that the trained algorithm selector is capable of selecting the most efficient algorithm that can cluster most accurately.

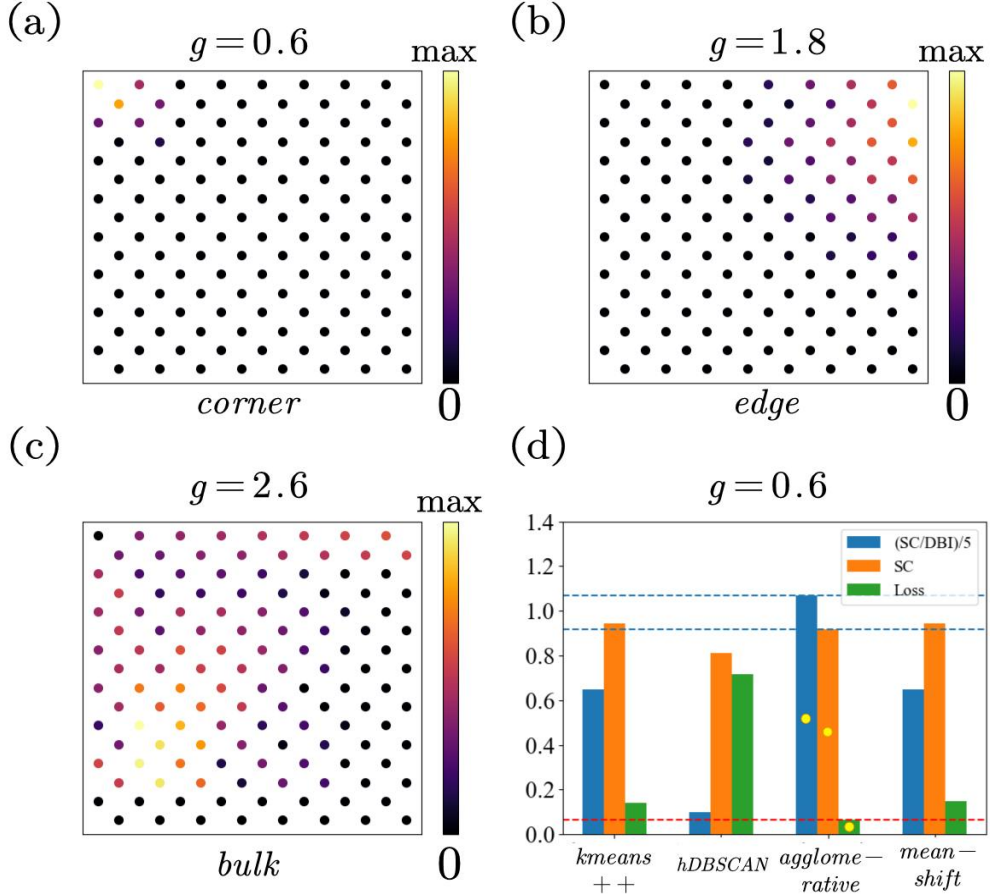

FIG. 5. Display of trained selector's clustering effect on the original model. (a)-(c) show the three kinds of topological modes for the original model with simple Floquet protocol. (a) displays corner mode with $g = 0.6$ (AFSOTI); (b) displays edge mode with $g = 1.8$ (AFTI); (c) displays bulk mode with $g = 2.6$ (NI). (d) is the comparison of the effectiveness of each algorithm in the algorithm library, and the selected one is marked with a yellow circle.

Thereafter, we employ the reliable selector on the fine-tuned model, especially on the singular topological phases AFFTTI and AFSTTI, and obtain the clusters containing eigenfunction indices. The eigenfunctions from different clusters are visualized in the form of eigenfield. Apart from conventional topological modes such as corner mode and edge mode, another two singular topological modes that integrate features of corner and edge mode, or edge and bulk mode, are disclosed. The singular topological modes are the manifestation of dynamic localization of the system, as the system is just in a "dynamic stable" state. The hybrid corner-edge mode and edge-bulk mode appear only under Floquet π mode with the ±π real part of quasi-energy and nonzero imaginary part. As the magnitude of gain/loss $g$ increases, the frequency of occurrence for both hybrid corner-edge mode and edge-bulk mode respectively go through a process of first increasing then decreasing, indicating that the system first enters dynamic stable Floquet mode, and then enters traditional stable Floquet mode again from the dynamic localization state.

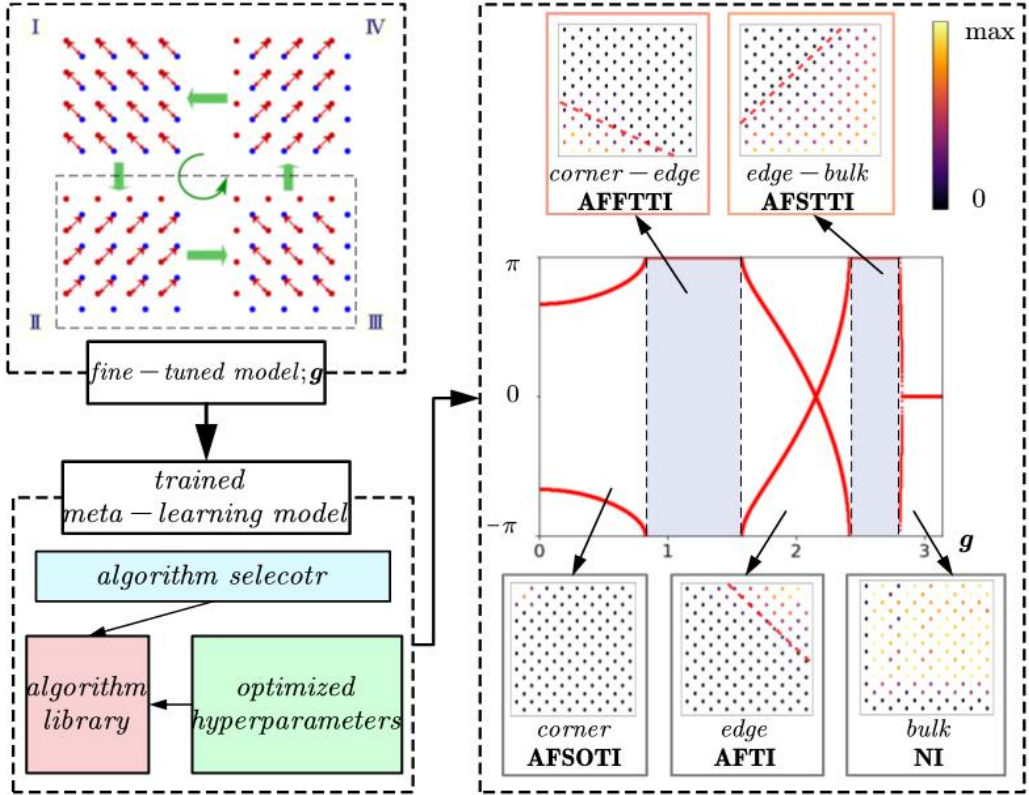

FIG. 6. Display of trained selector's clustering effect on the fine-tuned model. The right part shows eigenfields for five topological insulator mode that fine-tuned model system may support, including two kinds of transition mode. By drawing a prominent red dotted segment line, the singular eigenfunction distribution feature of hybrid topological mode can be visualized well, denoting the existence of dynamic localization stable.

The system enters five topological modes successively as magnitude of gain/loss increases, and the two transition Floquet π modes possess singular eigenfield distribution features, both of which integrates those of two adjacent modes. It means that the more complicated Floquet driving protocol will provide new possibilities of energy localization mode. Because of certain energy flow entailed by driving protocol, the fine-tuned model may not only rigidly localize at the corner or edge. Besides, it can dynamically localize at a hybrid corner-edge region. When the magnitude of gain/loss becomes larger, it can still localize in a little large regular region and not instantly tend to transform into the scattered bulk mode, supporting a hybrid edge-bulk mode. It should be pointed out that the topological mode transition is just a trend, meaning that the system does not only support one special topological mode at a certain value of $g$. When AFFTTI is supported, for instance, the eigenfield distribution may still show the form of corner mode, yet its probability of occurrence will consistently decrease as $g$ increases, meanwhile the edge mode will occur more frequently, until it almost entirely replaces the corner-edge mode, indicating that $g$ is large enough and the system begins to support next topological insulator AFTI.

## V. Conclusion

In this work, we introduce the unsupervised learning method and construct an algorithm selector model with a library containing a variety of clustering algorithms. One trained selector is utilized to test the basic model and explore the fine-tuned model, and eventually we discover the potential evolution regulation of the dynamic stabilization and localization of the system. Additionally, we build up a feasible methodological framework for solving such problems with machine learning method and clustering algorithms.

**Acknowledgements**
The authors thank for the support by National Natural Science Foundation of China under (Grant 12404365).

# References

[1] N. H. Lindner, G. Refael, and V. Galitski, Floquet topological insulator in semiconductor quantum wells, Nat. Phys. **7**, 490 (2011).

[2] T. Kitagawa, E. Berg, M. Rudner, and E. Demler, Topological characterization of periodically-driven quantum systems, Phys. Rev. B **82**, 235114 (2010).

[3] S. Yin, Emanuele Galiffi & Andrea Alù, Floquet metamaterials, Elight **2**,8 (2022).

[4] T. Oka and S. Kitamura, Floquet engineering of quantum materials, Annu. Rev. Conden. Ma. P. **10**, 387-408 (2019).

[5] Á. Sáiz, J. Khalouf-Rivera, J. M. Arias, P. Pérez-Fernández and J. Casado-Pascual, Quantum Phase Transitions in periodically quenched systems, Quantum **8**, 1365 (2024).

[6] T. Oka and H. Aoki, Photovoltaic hall effect in graphene, Phys. Rev. B **79**, 081406 (2009).

[7] T. Kitagawa, T. Oka, A. Brataas, L. Fu, and E. Demler, Transport properties of nonequilibrium systems under the application of light: Photoinduced quantum hallinsulators without landau levels, Phys. Rev. B **84**, 235108 (2011).

[8] Y. H. Wang, H. Steinberg, P. Jarillo-Herrero and N. Gedik, Observation of Floquet-Bloch states on the surface of a topological insulator, Science **342**, 453-457 (2013).

[9] G. E. Topp, G. Jotzu and J. W. McIver, Topological Floquet Engineering of Twisted Bilayer Graphene, Phys. Rev. Res. **1**, 023031 (2019).

[10] S. Rahav, I. Gilary, and S. Fishman, Time independent description of rapidly oscillating potentials, Phys. Rev. Lett. **91**, 110404 (2003).

[11] L. Li, C. H. Lee, and J. Gong, Realistic Floquet semimetal with exotic topological linkages between arbitrarily many nodal loops, Phys. Rev. Lett. **121**, 036401 (2018).

[12] L. Zhou, C. Chen, and J. Gong, Floquet semimetal with Floquet-band holonomy, Phys. Rev. B **94**, 075443 (2016).

[13] J. C. Budich, Y. Hu, and P. Zoller, Helical Floquet channels in 1d lattices, Phys.

Rev. Lett. **118**, 105302 (2017).

[14] A. F. Adiyatullin, L. K. Upreti, C. Lechevalier, C. Evain, F. Copie, P. Suret, S. Randoux, P. Delplace, and A. Amo, Topological properties of Floquet winding bands in a photonic lattice, Phys. Rev. Lett. **130**, 056901 (2023).

[15] M. S. Rudner, N. H. Lindner, E. Berg, and M. Levin, Anomalous edge states and the bulk-edge correspondence for periodically driven two-dimensional systems, Phys. Rev. X **3**, 031005 (2013).

[16] L. J. Maczewsky, J. M. Zeuner, S. Nolte, and A. Szameit, Observation of photonic anomalous Floquet topological insulators, Nat. Commun. **8**, 13756 (2017).

[17] B. Huang and W. V. Liu, Floquet higher-order topological insulators with anomalous dynamical polarization, Phys. Rev. Lett. **124**, 216601 (2020).

[18] L. Zhou and J. Gong, Recipe for creating an arbitrary number of Floquet chiral edge states, Phys. Rev. B **97**, 245430 (2018).

[19] K. Ding, C. Fang, and G. Ma, Non-Hermitian topology and exceptional-point geometries, Nat. Rev. Phys. **4**, 745 (2022).

[20] L. Zhou and D. Zhang, Non-Hermitian Floquet Topological Matter – A Review, Entropy **45**, 1401 (2023).

[21] H. Wang, X. Zhang, J. Hua, D. Lei, M. Lu, and Y. Chen, Topological physics of non-Hermitian optics and photonics: a review, J. Opt. **23**, 123001 (2021).

[22] C. Dembowski, H. D. Gräf, H. L. Harney, A. Heine, W. D. Heiss, H. Rehfeld, and A. Richter, Experimental observation of the topological structure of exceptional points, Phys. Rev. Lett. **86**, 787 (2001).

[23] S. K. Ozdemir, S. Rotter, F. Nori, and L. Yang, Parity–time symmetry and exceptional points in photonics, Nat. Mater. **18**, 783 (2019).

[24] M. A. Miri and A. Al`u, Exceptional points in optics and photonics, Science **363**, 6422 (2019).

[25] J. Doppler, A. A. Mailybaev, J. Böhm, U. Kuhl, A. Girschik, F. Libisch, T. J. Milburn, P. Rabl, N. Moiseyev, and S. Rotter, Dynamically encircling an exceptional point for asymmetric mode switching, Nature **537**, 76 (2016).

[26] L. L. C. H. L. Rijia Lin and T. Tai, Topological non-Hermitian skin effect, Fron. Phys. **18**, 53605 (2023).

[27] J. Pan and L. Zhou, Non-Hermitian Floquet second order topological insulators in periodically quenched lattices, Phys. Rev. B, **102**, 094305 (2020).

[28] X. Zhang, Y. Tian, J.-H. Jiang, M.-H. Lu, and Y.-F. Chen, Observation of higher-order non-Hermitian skin effect, Nat. Commun. **12**, 5377 (2021).

[29] L. Zhou and J. Gong, Non-Hermitian Floquet topological phases with arbitrarily many real-quasienergy edge states, Phys. Rev. B **98**, 205417 (2018).

[30] B. Höckendorf, A. Alvermann, and H. Fehske, Non-Hermitian boundary state engineering in anomalous Floquet topological insulators, Phys. Rev. Lett. **123**, 190403 (2019).

[31] S. Wu, W. Song, S. Gao, Y. Chen, S. Zhu, and T. Li, Floquet π mode engineering in non-Hermitian waveguide lattices, Phys. Rev. Res. **3**, 023211 (2021).

[32] L. Zhou, Non-Hermitian Floquet topological superconductors with multiple majorana edge modes, Phys. Rev. B **101**, 014306 (2020).

[33] S. Weidemann, M. Kremer, S. Longhi, and A. Szameit, Topological triple phase transition in non-Hermitian Floquet quasicrystals, Nature **601**, 354 (2022).

[34] L. Zhou, Y. Gu, and J. Gong, Dual topological characterization of non-Hermitian Floquet phases, Phys. Rev. B **103**, L041404 (2021).

[35] H. Gao, H. Xue, Z. Gu, L. Li, W. Zhu, Z. Su, J. Zhu, B. Zhang, and Y. D. Chong, Anomalous Floquet non-Hermitian skin effect in a ring resonator lattice, Phys. Rev. B 106, 134112 (2022).

[36] J. Gong, W. Zhu, L. Zhou and L. Li, Multiple topological transitions and spectral singularities in non-Hermitian Floquet, Phys. Rev. B **110**, 155413 (2024).

[37] L. Zhou, Dynamical characterization of non-Hermitian Floquet topological phases in one-dimension, Phys. Rev. B **100**, 184314 (2019).

[38] Q. Yan, B. Zhao, R. Zhou, R. Ma, Q. Lyu, S. Chu, X. Hu, Q. Gong, Advances and applications on non-Hermitian topological photonics, Nanophotonics **12**, 13 (2023).

[39] Y. Wu, B. Zhu, S. Hu, Z. Zhou and H. Zhong, Floquet control the gain and loss of PT symmetric optical couplers, Front. Phys. **12**, 121102 (2017).

[40] B. Wu, R. B. Diener, and Q. Niu, Bloch Waves and Bloch Bands of Bose-Einstein Condensates in Optical Lattices, Phys. Rev. A **65**, 025601 (2002).

[41] M. Bitter and V. Milner, Experimental observation of dynamical localization in laser-kicked molecular rotors, Phys. Rev. Lett. **117**, 144104 (2016).

[42] D. T. Liu, J. Shabani, and A. Mitra, Floquet Majorana zero and π modes in planar Josephson junctions, Phys. Rev. B **99**, 094303 (2019).

[43] Dunjko and Vedran, Quantum-Enhanced Machine Learning, Phys. Rev. Lett. 117,130501 (2016).

[44] G. Janzen, X. L. J. A. Smeets, V. E. Debets, C. Luo, C. Storm, L. M. C. Janssen and S. Ciarella, Dead or alive: Distinguishing active from passive particles using supervised learning, Europhys. Lett **143**, 17004 (2023).

[45] E. Bedolla, L. C. Padierna, R. Castañeda-Priego, Machine learning for condensed matter physics, J. Phys. Condens. Matter **33,** 053001 (2020).

[46] J. Lee, Y. Lee and Y. Whye, The Deep Amortized Clustering, arXiv: Learning, doi: 10.48550/ARXIV. 1909.13433. (2019).

[47] T. Hospedales, A. Antoniou, P. Micaelli and A. Storkey, Meta-Learning in Neural Networks: A Survey, IEEE Trans. Pattern Anal. Mach. Intell. **44**, 5149-5169 (2022).

[48] C. Finn, P. Abbeel and S. Levine, Model-Agnostic Meta-Learning for Fast Adaptation of Deep Networks, international conference on machine learning **70**, 1126-1135 (2017).

[49] A. Rivolli, L. P. F. Garcia, C. Soares, J. Vanschoren and A. C. P. L. F. de Carvalho, Meta-features for meta-learning, Knowl-Based Syst. **240**, 108101 (2022).

[50] Q. Wang, Y. Ma, K. Zhao and Y. Tian, A Comprehensive Survey of Loss Functions in Machine Learning, Ann. Data. Sci. **9**, 187–212 (2022).

[51] A. Chefrour and L. S. Meslati, Unsupervised Deep Learning: Taxonomy and Algorithms, Informatica **46**, 2 (2022).

[52] M. H. Mozaffari and S. H. Zahiri, Unsupervised data and histogram clustering

using inclined planes system optimization algorithm, Image Anal. Stereol. **33**, 65-74 (2014).

[53] Z. Wang and W. Shen, Analysis and identification of the compositional correlations of glasses and the variability of different classes of compositions, Acad. J. Mater. Chem. **4**, 5 (2023).

[54] J. Rousseeuw, Silhouettes: A graphical aid to the interpretation and validation of cluster analysis, J. Comput. Appl. Math. **20**, 53-65 (1985).

[55] D. T. Dinh, T. Fujinami and V. N. Huynh, Estimating the Optimal Number of Clusters in Categorical Data Clustering by Silhouette Coefficient, international symposium knowledge and systems sciences **1103**, 1-17 (2019).

# APPENDIX A: Meta-learning Core Idea and Basic Construction for Our Model

Meta-learning, also known as 'learning to learn', is a subfield of machine learning that shifts the focus from solving individual tasks to acquiring learning strategies capable of rapid adaptation to new, unseen tasks[1]. Unlike traditional machine learning, which optimizes model parameters for a single task, meta-learning seeks to discover meta-knowledge, such as parameter initialization, optimization rules, or architectural priors that enables efficient learning across a distribution of tasks. This approach is inspired by human cognition, where prior experience accelerates the acquisition of new skills.

The further analyzation for fine-tuned Floquet model in 'Amortized Clustering Assistant Classification of Anomalous Hybrid Floquet Modes in a Periodically Driven non-Hermitian Lattice' utilizes the core idea of meta-learning, and we originally build a basic meta-learning construction shown as Fig. S1.

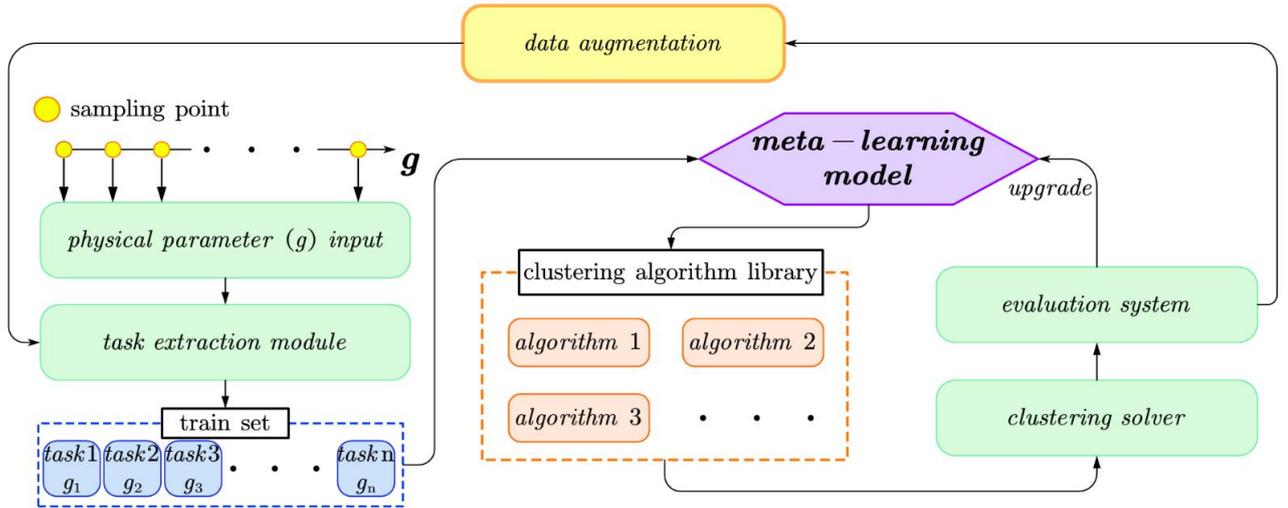

FIG. S1. Simplified framework of the machine learning method. The core section is the meta-learning model which is able to upgrade itself dynamically as the training progresses. The meta-learning model contains an algorithm library which the clustering solver select algorithm from. There is also a significant evaluation system that helps update the model and conduct data augmentation.

We take 300 different *g* values as physical parameter input. The task extraction module is able to produce multiple train tasks for the meta-learning model, whose core section is the clustering solver, i.e., an algorithm selector. Its outcomes will be sent to evaluation system, which calculates "loss function" and helps optimize the algorithm selection and hyper-parameter determination.

# APPENDIX B: Data Preprocessing and Visualization

Before the meta-learning model start to train formally, the dataset of each task, namely a variety of eigenfunctions, is supposed to be preprocessed in order to accelerate operating speed and avoid potential problems. In our work, the most vital process is the data standardization. Considering our dataset is high-dimensional and non-normal distribution, robust scaler is utilized to reduce the impact caused by outliers.

Robust scaler is a data preprocessing technique designed to enhance the robustness of feature scaling in machine learning pipelines, particularly in the presence of outliers [2]. Unlike traditional scaling methods such as standard scaler (which relies on mean and standard deviation) or min-max scaler (which uses minimum and maximum values), robust scaler employs median and interquartile range (IQR) for scaling.

The transformation formula for robust scaler is

$$x' = \frac{x - \overline{X}}{IQR(X)},$$

where $\overline{X}$ is the median of the data, and $IQR(X) = Q_3 - Q_1$, namely the difference between 75 and 25 quantile.

The key advantage of robust scaler is the outlier resistance, because median and IQR are less sensitive to extreme values. In addition, it can maintain the relative distribution of non-outlier data points, which is critical for models relying on feature relationships. Employing robust scaler on our raw high-dimensional dataset, the speed and reliability of follow-up work are guaranteed.

In order to visualize the high-dimensional data as well as clustering outcomes, we also have adopted t-Distributed Stochastic Neighbor Embedding (t-SNE) method [3]. It is a non-linear dimensionality reduction technique designed for visualizing high-dimensional data in low-dimensional spaces (typically 2D or 3D). Unlike linear methods such as PCA, t-SNE focuses on preserving local neighborhood relationships between data points, making it particularly effective for revealing clusters and non-linear structures in complex datasets. t-SNE operates by converting high-dimensional Euclidean distances between points into conditional probabilities that represent neighborhood relationships. The algorithm then minimizes the divergence between these probabilities and their low-dimensional counterparts using a Student's t-distribution (with heavy tails) to mitigate the "crowding problem" (where distant points in high dimensions collapse into a single cluster in low dimensions).

Firstly, neighborhood probability in high dimensional-space is calculated. For each data point $x_i$, the conditional probability of its neighbor $x_j$ is

$$p_{j|i} = \frac{\exp(-\|x_i - x_j\|^2 / 2\sigma_i^2)}{\sum_{k \neq i} \exp(-\|x_i - x_k\|^2 / 2\sigma_i^2)},$$

where $\sigma_i$ is determined via perplexity, which is a hyper-parameter controlling neighborhood size. The joint probability is defined as

$$p_{ij} = \frac{p_{j|i} + p_{i|j}}{2n},$$

where $n$ is the number of data points.

Secondly, in the embedded space, for each data point $y_i$, the similarity is modeled using a Student's t-distribution

$$q_{ij} = \frac{(1 + \|y_i - y_j\|)^{-1}}{\sum_{k \neq l} (1 + \|y_k - y_l\|)^{-1}}.$$

The optimization goal is to minimize the KL divergence:

$$KL = \sum_{i \neq j} p_{ij} \log \frac{p_{ij}}{q_{ij}}.$$

t-SNE excels at preserving local structure, revealing clusters and manifolds. It is robust to noise and outliers due to heavy-tailed t-distribution and widely applicable in exploratory data analysis (EDA) and visualization. By coloring the points corresponding to different clustering labels with different colors, t-SNE is able to display the clustering results. The Fig. S2 below show the clustering outcomes under several $g$ values for basic and fine-tuned model.

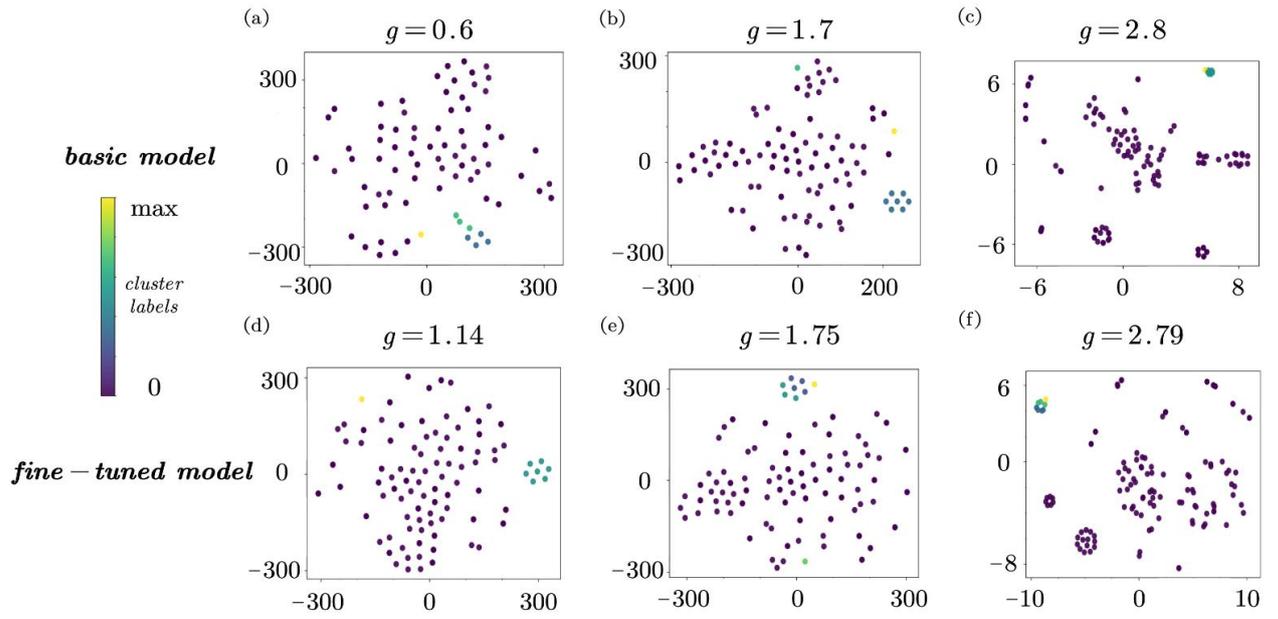

FIG. S2. Clustering results visualization in two-dimension image by t-SNE for several *g* values of basic and fine-tuned model. The clustering algorithms are all selected by the trained meta-learning model. (a)-(c) are for basic model, and (d)-(f) are for fine-tuned model.

# APPENDIX C: Algorithm Library and Hyper-parameter Optimization

The clustering algorithm library we build includes:

**1. K-means++**

The K-means++ algorithm is an improved K-means clustering algorithm [4-6]. By improving the selection strategy of initial cluster centers, the problems that may arise from randomly selecting initial cluster centers in traditional K-means algorithms have been avoided. The core idea is to keep the initial cluster centers as far apart as possible, thereby reducing the risk of the algorithm getting stuck in local optima.

Main hyper-parameter: number of clusters ($k$).

**2. hDBSCAN**

hDBSCAN is a hierarchical density-based clustering algorithm designed to address limitations of traditional methods like DBSCAN in handling variable-density clusters and automatic parameter selection [7-9]. By integrating hierarchical clustering with density-based spatial partitioning, hDBSCAN achieves robustness in complex datasets without requiring a predefined global density threshold (e.g., DBSCAN's ε parameter). Its core innovation lies in transforming density estimation into a hierarchical structure and extracting stable clusters through stability-based optimization.

Main hyper-parameter: min-cluster-size.

**3. agglomeration clustering**

The agglomerative clustering algorithm is a hierarchical clustering method that follows a bottom-up approach [10]. It starts by treating each data point as an individual cluster and iteratively merges the closest pairs of clusters until all points are clustered into one single cluster or a predefined stopping criterion is met. The distance between clusters can be measured using various linkage criteria, such as single-linkage, complete-linkage, or average-linkage. This method is commonly used in unsupervised learning tasks for discovering natural groupings within a dataset, and it is particularly useful when the number of clusters is not known beforehand. However, it can be computationally intensive for large datasets and sensitive to noise and outliers.

Main hyper-parameter: number of clusters, metric and linkage.

**4.mean-shift**

The mean-shift algorithm is an iterative algorithm based on kernel density estimation [11, 12]. Its core idea is to iteratively move data points to the location with the highest local density. Bandwidth is a vital parameter that determines the smoothness of kernel density estimation, which has a significant impact on clustering influence. In order to obtain a suitable bandwidth, Scott Law or Silverman Law may be considered, which requires standard deviation or interquartile range of the dataset. Or we can adopt estimate bandwidth function in python that allows automatic bandwidth estimation. The relative algorithm acquiring a suitable bandwidth may contribute to more operation time, yet the clustering outcomes considerably get better.

Main hyper-parameter: bandwidth.

We choose these classic algorithms and employ them on different tasks with a variety model input parameters $g$. Fig. S3 shows the Silhouette Coefficient (SC) [13] evolution with $g$ before (the upper section) and after (the lower section) hyper-parameters optimization. SC, as a profit index, represents better clustering outcomes with larger value. It is easily noted that the hDBSCAN algorithm possesses the worst clustering effect, and it is also rather instable.

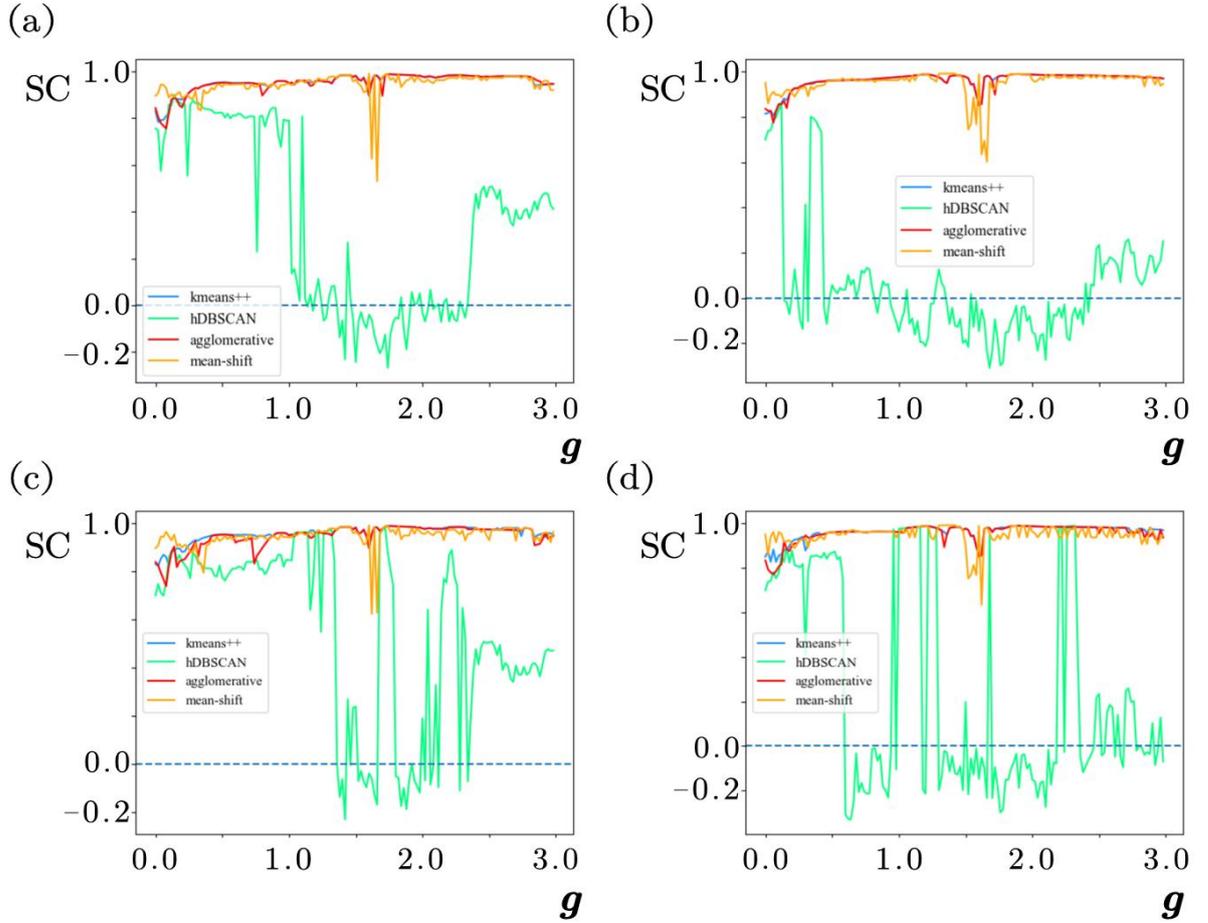

FIG. S3. SC evolution with g for four algorithms in the library. (a), (c) are for basic model; (b), (d) are for fine-tuned model. (a), (b) are plotted without any hyper-parameter optimization, with empirical default hyper-parameters. (c), (d) are plotted after hyper-parameter optimization train.

The hyper-parameter optimization adopts grid research, which is a systematic exhaustive search method employed for hyper-parameter optimization in machine learning and statistical modeling [14]. It operates by evaluating a predefined set of hyper-parameter configurations, often referred to as the parameter grid, through a structured exploration of the hyper-parameter space. The objective is to identify the combination of hyper-parameters that maximizes a specified performance metric (e.g., accuracy, F1-score, or mean squared error) on a validation set or via cross-validation.

Traditional grid research is able to find the global optimal solution in theory as long as the grid is dense enough, but at the cost of long operation time. We adopt random research [15], which samples configurations uniformly at random from the hyper-parameter space, offering better scalability for high-dimensional problems. The final optimization goal in our work is to minimize the value of "loss function", and the increase of SC is the direct index that indicates the algorithm selector is capable of clarifying the dataset better.

# References

[1] T. Hospedales, A. Antoniou, P. Micaelli and A. Storkey, Meta-Learning in Neural Networks: A Survey, IEEE Trans. Pattern Anal. Mach. Intell. **44**, 5149-5169 (2022).

[2] F. Pedregosa, G. Varoquaux, A. Gramfort, V. Michel, B.Thirion, O. Grisel, M.Blondel, P. Prettenhofer, R. Weiss, V. Dubourg, J. Vanderplas, A. Passos, D. Cournapeau, M. Brucher, M. Perrot and E. Duchesnay, Scikit-learn: Machine Learning in Python, J. Mach. Learn. Res. **12**, 2825-2830 (2011).

[3] L. Maaten and G. Hinton, Visualizing Data using t-SNE, J. Mach. Learn. Res. **9**, 2579-2605 (2008).

[4] J. Hämäläinen, T. Kärkkäinen and T. Rossi, Improving Scalable K-Means++, Algorithms **14**, 6 (2020).

[5] S. Yu and C. Yoon, Performance Improvement of Clustering Method Based on Random Swap Algorithm, Int. J. Fuzzy Log. Intell. Syst. **19**, 97-102 (2019).

[6] E. U. Oti, M.O. Olusola, F. C. Eze, and S. U. Enogwe, Comprehensive Review of K-Means Clustering Algorithms, Int. J. Adv. Sci. Res. Eng. **7**, 8 (2021).

[7] G. Stewart and M. Al-khassaweneh, An Implementation of the HDBSCAN Clustering Algorithm, Appl. Sci. **12**, 2405 (2022).

[8] S. Weng, J. Gou and Z. Fan, h-DBSCAN: A simple fast DBSCAN algorithm for big data, Proc. Mach. Learn. Res. **157**, 81-96 (2021).

[9] A. C. A. Neto, J. Sander, R. J. G. B. Campello, and M. A. Nascimento, Efficient Computation and Visualization of Multiple Density-Based Clustering Hierarchies, IEEE Trans. Knowl. Data Eng. **33**, 8 (2019).

[10] A. Hussain and P. Musilek, Applications of Clustering Methods for Different

Aspects of Electric Vehicles, Masooma Nazari, Electronics **12**, 4 (2023).

[11] D. Demirović, An Implementation of the Mean Shift Algorithm, Image Processing On Line **9**, 251-268 (2019).

[12] D. Comaniciu and P. Meer, Mean Shift: A Robust Approach Toward Feature Space Analysis, IEEE Trans. Pattern Anal. Mach. Intell. **24**, 5 (2022).

[13] J. Rousseeuw, Silhouettes: A graphical aid to the interpretation and validation of cluster analysis, J. Comput. Appl. Math. 20, 53-65 (1985).

[14] M. Adnan, A. A. S. Alarood, M. I. Uddin and I. ur Rehman, Utilizing grid search cross-validation with adaptive boosting for augmenting performance of machine learning models, PeerJ Comput. Sci. **8**, e803 (2022).

[15] Bengio, Yoshua, and J. Bergstra, Random Search for Hyper-Parameter Optimization, J. Mach. Learn. Res. **13**, 281–305 (2012).